\documentclass[prd,twocolumn,showpacs,amsmath,amssymb]{revtex4}

\usepackage{amssymb}
\usepackage{mathrsfs}
\usepackage{txfonts}

\usepackage{graphicx}
\usepackage{dcolumn}
\usepackage{bm}

\begin{document}
\title{Noncanonical warm inflation: a model with a general Lagrangian density}

\author{Kai Li}
\email{kai.li@qut.edu.cn}
\affiliation{School of Sciences, Qingdao University of Technology, Qingdao 266033, China}
\author{Xiao-Min Zhang}
\thanks{Corresponding author}
\email{zhangxm@mail.bnu.edu.cn}
\affiliation{School of Sciences, Qingdao University of Technology, Qingdao 266033, China}
\author{Hong-Yang Ma}
\email{hongyang_ma@aliyun.com}
\affiliation{School of Sciences, Qingdao University of Technology, Qingdao 266033, China}
\author{Jian-Yang Zhu}
\email{zhujy@bnu.edu.cn}
\affiliation{Department of Physics, Beijing Normal University, Beijing 100875, China}
\date{\today}

\begin{abstract}
Warm inflation is extended from the canonical field to the noncanonical field with a general Lagrangian density having coupling form of kinetic and potential terms. We develop the motion equation of the noncanonical inflaton field, and notice the motion equation has an annoying coupling term $\mathcal{L}_{X\varphi}\dot\varphi^2$, which makes the inflation difficult to solve. It is found that a special field redefinition exists, making the motion equation of inflaton uncoupled. Then we can solve inflation in the special field representation for convenience. The relation of field redefinition between the special uncoupled field and general noncanonical field is obtained. The motion equations and slow roll approximations of noncanonical warm inflation in the special field representation are developed. Then we give two examples to show how to translate a general noncanonical Lagrangian density in the coupled $\varphi$ representation, to the uncoupled $\phi$ representation. Finally, the cosmological perturbations generated by the new kind of warm inflation is calculated.

\end{abstract}
\maketitle

\section{\label{sec1}Introduction}
Inflation, a quasi-exponential expansion of the very early Universe, is a required supplement to the Big Bang Universe, which can successfully solve the problems of horizon, flatness and monopole \cite{Guth1981,Linde1982,Albrecht1982}. Inflation is also able to predict the generation of a nearly flat spectrum of primordial perturbations, fitting the observations of cosmological microwave background (CMB) exactly. Till now, there are two kinds of inflationary theories: standard inflation and warm inflation. Warm inflation, firstly proposed by A. Berera \cite{BereraFang,Berera1995}, has been developed a lot in the past more than twenty years. Warm inflation inherits the advantages of successfully solving the horizon and flatness problems and naturally producing seeds to give rise to the large scale structure of the Universe \cite{BereraFang,Lisa2004,Berera2000}. But the origins of the perturbation in the two kinds of inflation are different. Cosmological perturbations naturally arise from vacuum fluctuations of quantum fields in standard inflation \cite{LiddleLyth,Bassett2006}, while thermal fluctuations in warm inflation \cite{BereraFang,Lisa2004,Berera2000}. In addition, warm inflation does not need a separate reheating regime, and it can go smoothly to the radiation-dominated phase. Warm inflation can cure ``$\eta$-problem" \cite{etaproblem,etaproblem1} and the overlarge amplitude of the inflaton suffered in standard inflation \cite{Berera2006,BereraIanRamos}. The strict slow roll conditions in standard inflation can be relaxed a lot in warm inflation \cite{Ian2008,Campo2010,ZhangZhu,Zhang2014}. What's more, warm inflation broadens the scope of inflation, and thus some models that were ruled out in standard inflation such as the quartic chaotic potential model can again be in very good agreement with the Planck results in warm inflationary scenario \cite{Sam2014}.

Warm inflation was often considered as caused by a canonical inflaton field whose energy is dominated by potential, and the inflaton slow roll down its potential. An added thermal damping term makes the slow roll in warm inflation more easily to be satisfied \cite{Ian2008,Campo2010,ZhangZhu,Zhang2014}. The original picture of warm inflation was phenomenological \cite{BereraFang,Berera1995,Linde1999}, and A. Berera etc try to give the microphysical mechanism of warm inflation from the view of first principle \cite{Berera1999,Berera1998}. Then many works \cite{Berera1999,Berera1998,MossXiong,BereraRamos,Hall2005} concentrated on the thermal damping effect and give the mechanism that makes the warm inflation realizable in the point of field theory and particle physics \cite{Berera1999,Berera1998,MossXiong,BereraRamos,Hall2005}. Basing on these researches, the dissipative coefficient in different cases are obtained \cite{Zhangyi2009,ZhangZhu,BereraIanRamos,MossXiong,Gil2013,Linde1999,BereraRamos,Hall2004,Berera1998,Kephart,Ian2008,
Campo2010}. The research of perturbations in warm inflation are developed a lot in the past twenty years \cite{BereraFang,Berera1995,Berera2000,BereraIanRamos,Lisa2004,Taylor2000}. The perturbation equation of inflaton in warm inflation is second order Langevin equation including a thermal stochastic force. The early papers concentrating on perturbation in warm inflation get the analytic form of power spectrum \cite{Lisa2004,Berera2000,Taylor2000,Chris2009}, which is enhanced by the thermal effect compared to standard inflation. For scalar power spectrum is fixed by observations, the amplitude of primordial gravitation waves is thus decreased. With in-depth study, it's found that the analytic form power spectrum can be obtained just in the temperature independent case (the dissipative coefficient doesn't depend on temperature) \cite{Lisa2004,Chris2009}. In the temperature dependent case, it's hardly to get an analytic power spectrum, and an approximate analytic power spectrum basing the numerical method can be obtained \cite{Chris2009}. When the dissipative coefficient has a positive power law dependence of temperature, the power spectrum presents a ``growing mode" \cite{Chris2009}. The issue of non-Gaussianity has also been analysed in different cases such as strong or weak dissipative warm inflation, and temperature independent or dependent warm inflation \cite{Gupta2006,Gupta2002}.

Warm inflation are almost used the canonical scalar field as inflaton in the previous researches. Tachyon and Dirac-Born-Infeld (DBI) warm inflation are specially researched in \cite{Herrera2006,Cai2011} respectively. We proposed noncanonical warm inflation where the inflaton having an uncoupled Lagrangian density \cite{Zhang2014}. We find noncanonical warm inflation has some novel features such as more relaxed slow roll conditions and enhanced scalar power spectrum compared to canonical warm inflation. It's necessary and interesting to research noncanonical warm inflation, and noncanonical warm inflation broads the scope of inflationary theory. As for noncanonical warm inflation, the case we researched previously in \cite{Zhang2014} is just a simple kind to some degree. After the opening research for noncanonical warm inflation, we'll extend warm inflation to the more general case. In this paper, we concentrate on the noncanonical warm inflation where the Lagrangian density of inflaton having a coupled form of kinetic and potential terms. The coupled case is more general but complicated. The inflaton in the coupled noncanonical warm inflationary case may not have the ``right" or normal mass dimension as in canonical inflation. The emotion equation of inflaton in coupled noncanonical warm inflation is quite different from the uncoupled noncanonical warm inflation, let alone canonical warm inflation. In this paper, we establish the framework of coupled noncanonical warm inflation, try to get special field redefinition making the emotion equation of inflaton decoupled, and give the field redefinition relation between the special uncoupled field and the general noncanonical field. We develop the emotion equations and slow roll approximations of the coupled noncanonical warm inflation. Finally, we calculate the cosmological perturbations generated by the new kind of warm inflation.

The paper is organized as follows: In Sec. \ref{sec2}, we introduce the new noncanonical warm inflationary scenario, give the basic equations of the new picture, and obtain the relation between two different field representations. Then in Sec. \ref{sec3}, two concrete examples are given to show how transform an inflaton with a Lagrangian density having coupled form of kinetic and potential terms, to the inflaton with a Lagrangian density having uncoupled form. The perturbation calculations in the new scenario are performed in Sec. \ref{sec4}. Finally, we draw the conclusions and discussions in Sec. \ref{sec5}.

\section{\label{sec2}noncanonical warm inflation and field redefinition}
Warm inflation occurs when there is a significant amount of radiation production during the inflationary epoch, thus the Universe is hot with a non-zero temperature $T$.
In warm inflation, the Universe is a multi-component system whose total matter action is given by
\begin{equation}\label{action}
  S=\int d^4x \sqrt{-g}  \left[\mathcal{L}(X',\varphi)+\mathcal{L}_R+\mathcal{L}_{int}\right],
\end{equation}
where $X'=\frac12g^{\mu\nu}\partial_{\mu}\varphi\partial_{\nu}\varphi$. The Lagrangian density of the noncanonical field is $\mathcal{L}_{non-can}= \mathcal{L}(X',\varphi)$, which is potential dominated as usual, $\mathcal{L}_R$ is the Lagrangian density of radiation fields and $\mathcal{L}_{int}$ denotes the interactions between the scalar fields in warm inflation. Usually a proper noncanonical Lagrangian density should satisfy the conditions: $\mathcal{L}_{X'}\geq0$ and $\mathcal{L}_{X'X'}\geq0$ \cite{Franche2010,Bean2008}, where a subscript $X'$ denotes a derivative.

The emotion equation of inflaton can be obtained through varying the action:
\begin{equation}\label{vary}
  \frac{\partial(\mathcal{L}(X',\varphi)+\mathcal{L}_{int})}{\partial\varphi}-\left(\frac{1}{\sqrt{-g}}\right)
  \partial_{\mu}\left(\sqrt{-g}\frac{\partial\mathcal{L}(X',\varphi)}{\partial(\partial_{\mu}\varphi)}\right)=0.
\end{equation}
In the flat Friedmann-Robertson-Walker (FRW) Universe, the field is homogeneous, i.e. $\varphi=\varphi(t)$, and we can get the motion equation of the scalar field:
\begin{eqnarray}\label{EOM1}
 \left[\left(\frac{\partial\mathcal{L}(X',\varphi)}{\partial X'}\right)+2X'\left(\frac{\partial^2\mathcal{L}(X',\varphi)}{\partial X'^2}\right)\right]\ddot\varphi\nonumber\\+\left[3H\left(\frac{\partial\mathcal{L}(X',\varphi)}{\partial X'}\right)+\dot\varphi\left(\frac{\partial^2\mathcal{L}(X',\varphi)}{\partial X'\partial\varphi}\right)\right]\dot\varphi\nonumber\\-\frac{\partial(\mathcal{L}(X',\varphi)
 +\mathcal{L}_{int})}{\partial\varphi}=0,
\end{eqnarray}
where $X'=\frac12\dot\varphi^2$, and $H$ is the Hubble parameter. The thermal damping effect in warm inflation, which can be described by a dissipative coefficient $\Gamma$, comes from the interaction term $\mathcal{L}_{int}$ between inflaton and other sub-dominated scalar fields.

The energy-momentum tensor of the inflaton is $T^{\mu\nu}=\frac{\partial\mathcal{L}}{\partial X'}\partial^{\mu}\varphi\partial^{\nu}\varphi-g^{\mu\nu}\mathcal{L}$, and from that we can get the energy density and pressure of the inflaton: $\rho(\varphi,X')=2X'\frac{\partial\mathcal{L}}{\partial X'}-\mathcal{L}$, $p(\varphi,X')=\mathcal{L}$. An important characteristic parameter of noncanonical field which can describe the traveling speed of scalar perturbations is the sound speed:
\begin{equation}\label{soundspeed}
  c_s^2=\frac{\partial p/\partial X'}{\partial\rho/\partial X'}=\left(1+2X'\frac{\mathcal{L}_{X'X'}}{\mathcal{L}_X'}\right)^{-1},
\end{equation}
where the subscript $X'$ also denotes a derivative.

The Lagrangian density term $\mathcal{L}_{int}$ in Eq. (\ref{EOM1}) is just the function of zero order of the inflaton and other subdominated fields, but doesn't contain the derivative of the fields. The most successful explanation of the interaction between the inflaton and other fields is the supersymmetric two-stage mechanism \cite{MossXiong,Kephart}. Under some detailed calculations \cite{Berera1999,MossXiong}, the last term in Eq. (\ref{EOM1}), i.e. $\frac{\partial(\mathcal{L}(X',\varphi)+\mathcal{L}_{int})}{\partial\varphi}$ can be divided into two parts. One part expressed by $\tilde{\Gamma}\dot\varphi$ describes the dissipation effect of inflaton to all other fields \cite{BereraFang,Berera1995,Berera2000,Berera2006,BereraIanRamos}, another part is the sum of those terms which do not contain $\dot\varphi$ in the $\frac{\partial\mathcal{L}_{int}}{\partial\varphi}$ and the term $\frac{\partial\mathcal{L}(X',\varphi)}{\partial\varphi}$ in Eq. (\ref{EOM1}). The second part is resumed as the effective potential $V_{eff}$ in warm inflation, which is the potential acquired thermal correction and is a function of both inflaton and temperature \cite{BereraFang,Berera1999,MossXiong}. The temperature $T$ appearing in the effective potential is temperature of the radiation bath and does not fall to zero thanks to the dissipations of the inflaton to the bath \cite{BereraFang,Ian2008,MossXiong}. With these conventional calculations in warm inflation and using the definition of sound speed, the motion equation of the noncanonical inflaton can be expressed as:
\begin{equation}\label{EOM2}
  \mathcal{L}_{X'}c_{s}^{-2}\ddot{\varphi}+(3H\mathcal{L}_{X'}+\tilde{\Gamma})\dot{\varphi}+
  \mathcal{L}_{X'\varphi}\dot\varphi^2+V_{eff,\varphi}(\varphi,T)=0.
\end{equation}
The subscripts $\varphi$ and $X'$ both denote a derivative in this paper, and $\tilde{\Gamma}$ denotes the dissipative coefficient in warm inflation. From the equation above we can see that the damping terms, which contain an enhanced Hubble damping term and a thermal damping term, are much larger than that in the canonical warm inflation, let alone standard inflation. While, there is an annoying quadratic term of $\dot\varphi$ due to the kinetic potential coupling term of Lagrangian density $\mathcal{L}_{X'\varphi}$. The coupling term $\mathcal{L}_{X'\varphi}\dot\varphi^2$ brings difficulty to solve the inflation such as giving slow roll approximation and calculating perturbations to some extent. Fortunately, we can eliminate this term by making a field redefinition $\phi=f(\varphi)$. Using the new field representation, the Lagrangian density is $\mathcal{L}(X,\phi)$, where $X=\frac12\dot\phi^2$ in the FRW Universe. Then through varying the new action, we can get the motion equation of $\phi$, which of course has the same original form as Eq. (\ref{EOM2}):
\begin{equation}\label{EOM3}
  \mathcal{L}_{X}c_{s}^{-2}\ddot{\phi}+(3H\mathcal{L}_{X}+\Gamma)\dot{\phi}+
  \mathcal{L}_{X\phi}\dot\phi^2+V_{eff,\phi}(\phi,T)=0.
\end{equation}
This implies that we can often choose a field redefinition $\phi=f(\varphi)$ to make the coupling term vanish in the new representation by
\begin{equation}\label{coupling}
  \mathcal{L}_{X\phi}=\frac{1}{f^4_{\varphi}}[f_{\varphi}\mathcal{L}_{X'\varphi}-
  2f_{\varphi\varphi}\mathcal{L}_X'-2f_{\varphi\varphi}\mathcal{L}_{X'X'}X']=0,
\end{equation}
where $f_{\varphi}$ is the first derivative of the function $f(\varphi)$, and $f_{\varphi\varphi}$ is the second derivative of $f(\varphi)$. The dissipative coefficients in the two field representations can have the relation $\tilde{\Gamma}=f_{\varphi}^2\Gamma$. In uncoupling noncanonical warm inflation, the slow roll approximation equation Eq. (\ref{EOM5}) is easily to be satisfied, which we'll discuss hereafter. So we can express the time derivative of $\phi$ in terms of $\phi$ as, say $\dot\phi=g(\phi)$, and also we have $X=\frac12g^2(\phi)$, $X'=\frac{g^2(f(\varphi))}{2f_{\varphi}^2}$. Then, theoretically, given a coupled form Lagrangian density $\mathcal{L}(X',\varphi)$, combining Eqs. (\ref{coupling}) and (\ref{EOM5}), we can work out the function $f_{\varphi}$ and then $f(\varphi)$ analytically or with the help of numerical method.

Thus when using the new redefined $\phi$ representation, the motion equation of inflaton becomes:
\begin{equation}\label{EOM4}
   \mathcal{L}_{X}c_{s}^{-2}\ddot{\phi}+(3H\mathcal{L}_{X}+\Gamma)\dot{\phi}+V_{eff,\phi}(\phi,T)=0.
\end{equation}
For simplicity we'll write the thermal effective potential $V_{eff}$ as $V$ hereinafter. Now the motion equation of inflaton has a clear form: a second order inertia term, a first order damping term and the effective potential term. Then the slow roll approximations, the warm inflationary dissipative strength and the perturbation quantities etc are more easily to be worked out, which we'll analyse hereafter. We'll give calculation of noncanonical warm inflation in the easy-to-use $\phi$ representation hereafter.

Given an arbitrary noncanonical Lagrangian density $\mathcal{L}(X',\varphi)$ or its transformed form Lagrangian density $\mathcal{L}(X,\phi)$, the noncanonical inflatons and the dissipative coefficient, may not has the normal mass dimension as in canonical warm inflation, so the dimensionless dissipative strength parameter in our noncanonical warm inflation is defined as
\begin{equation}\label{r}
  r=\frac{\Gamma}{3H\mathcal{L}_X},
\end{equation}
which is different from canonical warm inflation. Warm inflation is in strong regime when $r\gg1$, while in weak regime when $r\ll1$.

The thermal damping effect of inflaton and energy transfer to radiation in warm inflation \cite{BereraIanRamos,Zhang2014} can be characterized by the entropy production equation:
\begin{equation}\label{entropy1}
    T\dot{s}+3HTs=\Gamma\dot{\phi}^{2},
\end{equation}
where $s$ is the entropy density.

Inflation is often associated with slow roll approximations to drop the highest derivative terms in the equations of motion, the same in the new inflationary picture. The slow roll equations are:
\begin{equation}\label{EOM5}
   (3H\mathcal{L}_{X}+\Gamma)\dot\phi+V_{\phi}(\phi,T)=0,
\end{equation}
\begin{equation}\label{entropy}
    3HTs-\Gamma \dot\phi^2=0.
\end{equation}
The validity of slow roll approximations depends on the slow roll conditions \cite{Zhang2014}:
\begin{eqnarray}
\epsilon&\ll&\frac{\mathcal{L}_{X}(1+r)}{c^2_s},~~\beta\ll\frac{\mathcal{L}_{X}(1+r)}{c^2_s},~~\eta\ll\frac{\mathcal{L}_{X}}{c^2_s},
\nonumber\\ &b&\ll\frac{min\{1,r\}}{(1+r)c^2_s},~~~~~~~|c|<4.
\end{eqnarray}
The slow roll parameters in the equations above are defined as:
\begin{equation}
\epsilon =\frac{M_p^2}{2}\left(\frac{V_{\phi}}{V}\right) ^2, ~~~\eta =M_p^2\frac {V_{\phi \phi}}{V}, ~~~\beta
=M_p^2\frac{V_{\phi}\Gamma_{\phi}}{V\Gamma},
\end{equation}
another two additional parameters describing the temperature dependence are:
\begin{equation}
b=\frac {TV_{\phi T}}{V_{\phi}},~~~~c=\frac{T\Gamma_T}{\Gamma},
\end{equation}
where $M_p^2=1/8\pi G$ is the reduced squared Planck mass.
The slow roll approximations are more easily to be guaranteed in noncanonical warm inflationary scenario than in canonical warm inflation, let alone standard inflation, so we can safely use them in the calculation of Sec. \ref{sec4}.

The number of e-folds in the noncanonical warm inflation can be given by
\begin{equation}
N=\int H dt=\int\frac{H}{\dot{\phi}}d\phi\simeq-\frac{1}{M_p^2}\int_{\phi_{\ast}}
^{\phi_{end}}\frac{V\mathcal{L}_X(1+r)}{V_{\phi}}d\phi,
\end{equation}
where $\phi_{\ast}$ is the inflaton value when horizon crossing.

\section{\label{sec3}two examples}
In this section, we'll give two concrete examples to show how to translate a general noncanonical Lagrangian density with $\mathcal{L}_{X'\varphi}\neq0$ in the coupled $\varphi$ representation, to the easy-to-use uncoupled representation.

\subsection{\label{sec31} the model with $\mathcal{L}(X',\varphi)_{X'X'}=0$}
Given an original noncanonical Lagrangian density $\mathcal{L}(X',\varphi)=e^{-\varphi}X'-V_0 e^{-2\varphi}$, we can calculate its $\mathcal{L}_{X'}=e^{-\varphi}$, $\mathcal{L}_{X'\varphi}=-e^{-\varphi}\neq0$. Basing on the Eq. (\ref{coupling}), we should have:
\begin{equation}\label{coupling1}
 f_{\varphi}\mathcal{L}_{X'\varphi}-2f_{\varphi\varphi}\mathcal{L}_{X'}=0
\end{equation}
to make $\mathcal{L}_{X\phi}=0$. Then we can work out that if
\begin{equation}\label{f1}
  \phi=f(\varphi)=-2e^{-\frac12\varphi},
\end{equation}
the Lagrangian density can be expressed as
\begin{equation}\label{Lagrangian1}
  \mathcal{L}(X,\phi)=X-\frac1{16}V_0\phi^4,
\end{equation}
where we can see that the coupling term $\mathcal{L}_{X\phi}=0$.

\subsection{\label{sec32} the model with $\mathcal{L}(X',\varphi)_{X'X'}\neq0$}
We begin with a coupled form Lagrangian density which can be expressed as $\mathcal{L}(X',\varphi)=e^{-\varphi}\ln X'-V(\varphi)$, then $\mathcal{L}_{X'}=\frac{e^{-\varphi}}{X'}$, $\mathcal{L}_{X'X'}=-\frac{e^{-\varphi}}{X'^2}$, and $\mathcal{L}_{X'\varphi}=-\frac{e^{\varphi}}{X'}$. The coupling term of the motion equation Eq. (\ref{EOM2}) in $\varphi$ representation is
\begin{equation}\label{coupling2}
  \mathcal{L}_{X'\varphi}\dot\varphi^2=2\mathcal{L}_{X'\varphi}X'=-2e^{-\varphi}.
\end{equation}

From the equation above, we can see the coupling term is totaly a function of $\varphi$, but not of $\dot\varphi$, so it can be well absorbed into the effective dominated potential term $V_{eff}$. Then the motion equation of $\varphi$ still can be expressed in an uncoupling form.

\section{\label{sec4}perturbations in the noncanonical warm inflation}

Now we perform the theory of cosmological perturbations in the noncanonical warm inflation. Considering the small perturbations, we can expand the total inflaton field as $\Phi(\mathbf{x},t)=\phi(t)+\delta\phi(\mathbf{x},t)$, where $\delta\phi(x,t)$ is the linear response due to the thermal stochastic noise $\xi$ in thermal system. The thermal noise source we introduced in warm inflation is Markovian: $\langle\xi(\mathbf{k},t)\xi(-\mathbf{k}',t)\rangle=2\Gamma T(2\pi)^3\delta^3(\mathbf{k}-\mathbf{k}')\delta(t-t')$ \cite{Lisa2004,Gleiser1994}. Introducing the noise term and the dissipative term, and substituting the expansion of inflaton to the motion equation which obtained through varying the action, we can get a second order Langevin equation:
\begin{eqnarray}\label{perturbation1}
  \mathcal{L}_{X}c_s^{-2}(\ddot\phi(t)+\delta\ddot\phi(\mathbf{x},t))&+&(3H\mathcal{L}_X+\Gamma)(\dot\phi(t)+
  \delta\dot\phi(\mathbf{x},t))\nonumber\\+V_{\phi}+V_{\phi\phi}\delta\phi(\mathbf{x},t)&-&
  \mathcal{L}_X\frac{\nabla^2}{a^2}\delta\phi(\mathbf{x},t)=\xi(\mathbf{x},t).
\end{eqnarray}
After we taking the Fourier transform, the evolution equation for the fluctuation can be obtained:
\begin{equation}\label{perturbation2}
  \mathcal{L}_{X}c_s^{-2}\delta\ddot\phi_{\mathbf{k}}+(3H\mathcal{L}_X+\Gamma)\delta\dot\phi_{\mathbf{k}}+
  (\mathcal{L}_X\frac{k_c^2}{a^2}+m^2)\delta\phi_{\mathbf{k}}=\xi_{\mathbf{k}},
\end{equation}
where $m^2=V_{\phi\phi}$ and $m^2$ is a tiny term that much less than the term $\mathcal{L}_Xk_p^2$ in slow roll inflation (the relation between comoving wave number $k_c$ and physical wave number $k_p$ is $k_p=k_c/a$).

The second order Langevin equation above is hard to solve, and we only want to get the perturbation observations when horizon crossing. Horizon crossing is well inside the slow roll inflationary regime \cite{LiddleLyth}. The warm slow roll regime is overdamped so the inertia term can be neglected \cite{Berera2000,Taylor2000}. Then the second order Langevin equation (\ref{perturbation2}) can be reduced to first order:
\begin{equation}\label{perturbation3}
  3H\mathcal{L}_X(1+r)\delta\dot\phi_{\mathbf{k}}+
  (\mathcal{L}_X\frac{k_c^2}{a^2}+m^2)\delta\phi_{\mathbf{k}}=\xi_{\mathbf{k}}.
\end{equation}
The approximate solution is given by
\begin{eqnarray}\label{solution}
  \delta\phi_{\mathbf{k}}(\tau)\approx \frac1{3H\mathcal{L}_{X}(1+r)}\exp\left[-\frac{\mathcal{L}_Xk_p^2+m^2}
  {3H\mathcal{L}_X(1+r)}(\tau-\tau_0)\right]\nonumber \\ \int_{\tau_0}^{\tau}\exp\frac{\mathcal{L}_Xk_p^2+m^2}
  {3H\mathcal{L}_X(1+r)}(\tau'-\tau_0)\xi(\mathbf{k},\tau')d\tau' \nonumber \\ +
  \delta\phi_{\mathbf{k}}(\tau_0)\exp\left[-\frac{\mathcal{L}_Xk_p^2+m^2}
  {3H\mathcal{L}_X(1+r)}(\tau-\tau_0)\right].
\end{eqnarray}
With the solution, we can get the corresponding correlation function
\begin{eqnarray}\label{correlation}
  \langle\delta\phi_{\mathbf{k}_p}(\tau)\delta\phi_{\mathbf{k}_p'}(\tau)\rangle\approx (2\pi)^3
  \frac{\Gamma T}{3H\mathcal{L}_X(1+r)(\mathcal{L}_Xk_p^2+m^2)}
  \nonumber\\ \delta^3(\mathbf{k}_p-\mathbf{k}_p')\left[1-\exp\left(-\frac{2(\mathcal{L}_Xk_p^2+m^2)}{3H\mathcal{L}_X(1+r)}(\tau-\tau_0)\right)\right]
  \nonumber \\
  +\langle\delta\phi_{\mathbf{k}_p}(\tau_0)\delta\phi_{\mathbf{k}_p'}(\tau_0)\rangle \exp\left[-\frac
  {2(\mathcal{L}_Xk_p^2+m^2)}{3H\mathcal{L}_X(1+r)}(\tau-\tau_0)\right].
\end{eqnarray}
On the right hand side of Eq. (\ref{solution}), the first term is the noise contribution that is driving the mode to thermal equilibrium, and the second term contains the memory of the initial conditions at $\tau=\tau_0$, which are exponentially damped. In the expanding Universe, from Eq. (\ref{solution}), we can find that the larger $k_p^2$ is the faster the relaxation rate is. If $k^2_p$ is sufficiently large for the mode to relax within a Hubble time, then that mode thermalizes. As soon as the physical wave number of a $\delta\phi(\mathbf{x},t)$ field mode becomes less than the freezing wave number $k_F$, it essentially feels no effect of the thermal noise $\xi(\mathbf{k},t)$ during a Hubble time \cite{Berera2000}. Based on the criterion, the freeze-out physical wave number $k_F$ is defined as $\frac{\mathcal{L}_Xk_F^2+m^2}{3H\mathcal{L}_X(1+r)H}=1$. In slow roll inflation, the mass term is negligible, so we can work out the freeze-out wave number
\begin{equation}\label{freezeout}
  k_F=\sqrt{3(1+r)}H.
\end{equation}
The freezeout time $t=t_F$ always precedes the horizon crossing time, so the scalar power spectrum of warm inflation is already fixed at an early time $k_F$ when $k=k_F>H$. Through the relation $\langle\delta\phi_{\mathbf{k}}\delta\phi_{\mathbf{k}'}\rangle=\delta^3(\mathbf{k}-\mathbf{k}')\frac{2\pi^2}{k^3}
\mathcal{P}_{\phi}(k)$ and Eq. (\ref{correlation}), we can get the power spectrum of inflaton:
\begin{equation}\label{powerphi}
  \mathcal{P}_{\phi}(\mathbf{k})=\frac{rk_FT}{2\pi^2(1+r)\mathcal{L}_X}=\frac{\sqrt{3}rHT}{2\pi^2\sqrt{1+r}\mathcal{L}_X}.
\end{equation}
Then the scalar power spectrum of curvature perturbation is
\begin{equation}\label{power}
  \mathcal{P}_R=\left(\frac{H}{\dot\phi}\right)^2\mathcal{P}_{\phi}=\frac{9\sqrt{3}H^5\mathcal{L}_X(1+r)^{\frac32}rT}{2\pi^2V_{\phi}^2}.
\end{equation}

In strong regime of warm inflation ($r\gg1$), the power spectrum becomes
\begin{equation}\label{strong}
  \mathcal{P}_R=\left(\frac{H}{\dot\phi}\right)^2\mathcal{P}_{\phi}=\frac{9\sqrt{3}H^5\mathcal{L}_Xr^{\frac52}T}{2\pi^2V_{\phi}^2},
\end{equation}
and in weak regime of warm inflation ($r\ll1$), the power spectrum becomes
\begin{equation}\label{weak}
  \mathcal{P}_R=\left(\frac{H}{\dot\phi}\right)^2\mathcal{P}_{\phi}=\frac{9\sqrt{3}H^5\mathcal{L}_XrT}{2\pi^2V_{\phi}^2}.
\end{equation}

Cosmological microwave background (CMB) observations provide a good normalization of the scalar power spectrum $P_R\approx10^{-9}$ on large scales. We can see from our new result Eq. (\ref{power}) that, compared to standard inflation or canonical warm inflation, the energy scale when horizon crossing can be depressed by both the noncanonical effect and thermal effect. This is good news to the assumption that the universe inflation can be described well by effective field theory.

The spectral index is
\begin{equation}\label{index}
  n_s-1=\frac {d\ln\mathcal{P}_R} {d\ln k} \simeq \frac{  \mathcal{\dot P}_R} {H\mathcal{P}_R},
\end{equation}
which can be expressed by
\begin{eqnarray}\label{index1}
  n_s-1&=&\left[\frac{5c-16}{4-c}+\frac{6r}{(4-c)(1+r)}\right]\frac{\epsilon}{\mathcal{L}_X(1+r)}\nonumber\\
  &+&\frac{2\eta}{\mathcal{L}_X(1+r)}-\frac{10r+4}{(4-c)(1+r)}\frac{\beta}{\mathcal{L}_X(1+r)}\nonumber\\&-&
  \frac{3r}{2(1+r)}\left(\frac{1}{c_s^2}-1\right)\delta +\frac{5cr+2r+2c+2}{2(4-c)r}b\nonumber\\&+&
  \frac1{4-c}\frac{\epsilon\beta}{\mathcal{L}_X(1+r)}
\end{eqnarray}
in our noncanonical warm inflation, where $\delta=\frac{\ddot\phi}{H\dot\phi}$ is also a slow roll parameter which is much less than one. Through the equation above and the slow roll conditions in noncanonical warm inflation we stated in Sec. \ref{sec2}, we can see that the spectral index, which is of order $\frac{\epsilon}{\mathcal{L}_X(1+r)}$, is much less than one. So we obtained a nearly scale-invariant power spectrum that is consistent with observations in general noncanonical warm inflation.

Since the tensor perturbations do not couple to the thermal background, the gravitational waves are only generated by the quantum fluctuations as in standard inflation:
\begin{equation}\label{tensor}
  \mathcal{P}_T=\frac2{M_p^2}\left(\frac H{2\pi}\right)^2.
\end{equation}
The tensor-to-scalar ratio thus is
\begin{equation}\label{ratio}
  R=\frac{\mathcal{P}_T}{\mathcal{P}_R}=\frac HT\frac{2\epsilon}{\sqrt{3}\mathcal{L}_X(1+r)^{\frac32}r}.
\end{equation}
From the equation above, we can find that the tensor perturbations can be weaker than canonical warm inflation, let alone standard inflation. This characteristic is due to both the noncanonical effect and thermal effect if both the effects are strong, which is a kind of synergy of both effects.

\section{\label{sec5}conclusions and discussions}
We summarize with a few remarks. We extend warm inflation to more general noncanonical case and thus generalize the scope of the inflationary theory. The general warm inflation are often dominated by a noncanonical inflaton with a complicated Lagrangian density $\mathcal{L}(X',\varphi)$ having $\mathcal{L}_{X'\varphi}\neq0$. This kind noncanonical warm inflation is not easy to solve, for an annoying coupling term $\mathcal{L}_{X'\varphi}\dot\varphi^2$ presenting in the motion equation makes the slow roll inflation hardly to be defined and given out. Fortunately, in many cases, we can transform the Lagrangian density $\mathcal{L}(X'\varphi)$ which has an annoying coupling form to the clear uncoupling form $\mathcal{L}(X,\phi)$ with the help of a field redefinition $\phi=f(\varphi)$. Then in the new filed representation, the noncanonical Lagrangian density $\mathcal{L}(X,\phi)$ has $\mathcal{L}_{X\phi}=0$. Then the warm inflation can be more easily to be dealt, for the motion equation has an conventional uncoupling form, the slow roll approximations are clear to be defined and used. In noncanonical warm inflation, the slow roll conditions are more easily to be satisfied, for the Hubble damping term is enhanced by noncanonical effect and there is an additional thermal damping term. We give two concrete examples to show that how to make field redefinitions to obtain uncoupling noncanonical warm inflation in different cases.

In the new kind warm inflationary model, we calculate and obtain a new form but still nearly scale invariant scalar power spectrum. And we find the energy scale during horizon crossing can be depressed by the synergy of the noncanonical effect and thermal effect. The tensor-to-scalar ratio is also analysed, and we find the amplitude of gravitational wave is weaker than canonical warm inflation and standard inflation. This characteristic can result from both the noncanonical effect and thermal effect if both the effects are strong, which is another synergy of both effects. The detailed issue of non-Gaussianity in the new scenario also deserves more cognition and research, which can be our future work, and we will also concentrate on a new kind of warm inflation which is not dominated by potential again.

\acknowledgments This work was supported by the National Natural Science Foundation of China (Grants No. 11605100 and No. 11547035).

\end{document}